\documentclass[10pt,letterpaper]{article}
\usepackage{opex3}

\begin{document}

\title{Engineering of directional emission from photonic crystal waveguides}

\author{Steven K. Morrison and Yuri S. Kivshar}

\address{Nonlinear Physics Centre and Centre for Ultra-high
bandwidth Devices for Optical Systems (CUDOS), Research School of
Physical Sciences and Engineering, Australian National University,
Canberra, ACT 0200, Australia}

\email{skm124@rsphysse.anu.edu.au} 



\begin{abstract}
We analyze, by the finite-difference time-domain numerical
methods, several ways to enhance the directional emission from
photonic crystal waveguides through the beaming effect recently
predicted by Moreno {\em et al.} [Phys. Rev. E {\bf 69}, 121402(R)
(2004)], by engineering the surface modes and corrugation of the
photonic crystal surface. We demonstrate that the substantial
enhancement of the light emission can be achieved by {\em
increasing} the refractive index of the surface layer. We also
measure power of surface modes and reflected power and confirm
that the enhancement of the directional emission is related to the
manipulation of the photonic crystal surface modes.
\end{abstract}




One of the recent advances in the physics of photonic crystals is
the discovery of enhanced transmission and highly directional
emission from photonic crystal waveguides predicted theoretically
by Moreno {\em et al.}~\cite{moreno} and demonstrated
independently in experiment by Kramper {\em et al.}~\cite{costas}.
These results provide a new twist in the study of surface modes in
photonic crystals. Indeed, it is generally believed that surfaces
and surface modes are a highly undesirable feature of photonic
crystals, unlike point defects which are useful for creating
efficient waveguides with mini-band gaps inside the photonic band
gaps of a periodic structure. However, appropriate corrugation of
the surface layer may lead to coherent enhancement of the
radiating surface modes and highly directional emission of the
light from a truncated waveguide~\cite{moreno,costas}.

As already mentioned by Moreno {\em et al.}~\cite{moreno}, the
major motivation for the discovery of highly directional emission
from photonic crystal waveguides is largely provided by the
physics of extraordinary optical transmission through
subwavelength hole arrays in metallic thin films~\cite{ebbesen}
and beaming of light from single nanoscopic apertures franked by
periodic corrugations~\cite{lezec}. In both those cases, an
incident light beam couples to the surface plasmon oscillations
via corrugations in a metallic film, and is then emitted from the
other side of the film being enhanced by its other corrugated
surface. For photonic crystal waveguides, properties of the
surface layer~\cite{moreno} or terminated surface~\cite{costas}
provide a key physical mechanism for the excitation of surface
modes, their constructive interference, and subsequent highly
directed emission.

In this paper, we study, by means of the finite-difference
time-domain (FDTD) numerical method, the directional emission from
a photonic crystal waveguide achieved by appropriate corrugation
of the photonic crystal interface, following the original
suggestion~\cite{moreno}. We analyze several strategies for
enhancing the light beaming effect by varying the surface
properties and by engineering the surface modes of a semi-infinite
two-dimensional photonic crystal created by a square lattice of
cylinders in vacuum. In particular, we optimize the corrugation at
the surface, as well as vary the refractive index of the surface
layer. We demonstrate that, in comparison with the previously
published results~\cite{moreno}, the substantial enhancement of
the light emission and improved beaming effect can be achieved by
{\em increasing} the refractive index of the surface layer while
using a positive (i.e. opposite to that employed in
Ref.~\cite{moreno}) corrugation displacement. We also measure the
power of surface modes and reflected power and confirm that the
enhancement of the directional emission through the beaming effect
links closely to the manipulation of the surface modes supported
by the photonic crystal interface.

\begin{figure}[htbp]
\centering\includegraphics[width=10cm]{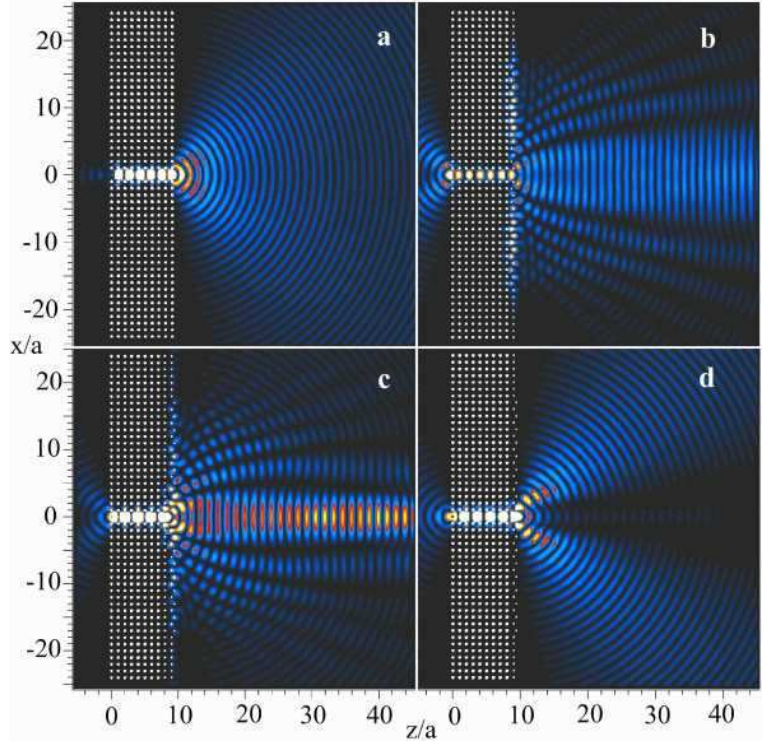}
\caption{Spatial distribution of the Poynting vector for the light
emitted from a photonic waveguide: (a) unchanged surface; (b)
surface cylinders with $r_{\rm s}=0.09$ and $N=9$ even-numbered
cylinders displaced by $\Delta z=-0.3 \, a$ (see~\cite{moreno});
(c) surface cylinders with $r_{\rm s}=0.09$, refractive index
$n_{s}=3.6$, and $N=9$ odd-numbered cylinders displaced by $\Delta
z=+0.4 \, a$; in addition, the radius of the cylinders in the
layer prior to the surface layer is reduced to $r_{\rm s-1}=0.135
\, a$; (d) surface cylinders with $r_{\rm s}=0.09$, refractive
index $n_{\rm s}=4.5$, and $N=9$ odd-numbered cylinders displaced
by $\Delta z=+0.4 \, a$.}\label{Poynting_2_Layers}
\end{figure}

We consider a photonic crystal slab created by a square lattice of
cylinders with dielectric constant $\epsilon_{r}=11.56$ (e.g. GaAs
at a wavelength of 1.5 $\mu$m) and radius $r=0.18 \, a$, where $a$
is the lattice period. A row of cylinders removed along the plane
$x=0$ forms a single-mode waveguide (see
Fig.~\ref{Poynting_2_Layers}) that supports a guided mode with
frequencies between $\omega =0.30 \times 2\pi c/a$ and $\omega =
0.44\times 2\pi c/a$ propagating in the plane normal to the
cylinders, with the electric field parallel to them.

When a source is placed in the waveguide at the point $z=0$, it
excites waves that propagate along the waveguide and are then
emitted at the waveguide exit (at $z =9a$). Since no surface modes
are supported by a simple truncated slab, the light radiating from
the waveguide undergoes uniform angular diffraction as
demonstrated in Fig.~\ref{Poynting_2_Layers}(a) for the spatial
distribution of the Poynting vector calculated for the source
frequency $\omega = 0.408 \times 2\pi c/a$.

To characterize the transmission from the photonic crystal
waveguide, we measure the directed power $P_{\rm D}$, normalized
to the input power, incident upon a cross-sectional length of $2a$
centered at $x=0$ and $z=45\,a$. A likewise normalized measure is
taken of the reflected power $P_{\rm R}$ incident upon a
cross-sectional length of $20 \, a$ centered at the input to the
waveguide, $x=0$ and $z=-a$. This reflected power is considered a
close measure of all reflected power. For the bulk photonic
crystal with standard surface layer the directed power is $P_{\rm
D}=0.0123$, and the reflected power is $P_{\rm R}=0.0158$.

Distribution of the Poynting vector for the directional emission
from the photonic crystal waveguide demonstrated by Moreno {\em et
al.}~\cite{moreno} is shown in Fig.~\ref{Poynting_2_Layers}(b).
These results are produced by altering the surface layer geometry
in two ways. Firstly, by reducing the radius of the surface
cylinders to the value $r_{\rm s}=0.5r=0.09\,a$, and thereby
creating the conditions for a surface mode to exist at the
truncated surface. And secondly, by displacing $N=9$ even-numbered
cylinders (numbered consecutively away from the waveguide) on both
sides of the waveguide by $\Delta z=-0.3\,a$ along the $z-$axis of
the crystal, thus enhancing radiation of surface modes. Our
calculations show that the directed power for such a structure is
$P_{\rm D}=0.0723$, while the reflected power is substantially
large, $P_{\rm R}=0.2635$. To further characterize the enhanced
beaming effect, we measure one half of the total surface mode
power, $P_{\rm S}$, incident upon a cross-sectional length $2\,a$
positioned centrally at $x=24\,a$, $z=9\,a$; again normalized to
the input power. Moreover, to characterize the containment of the
directed power we measure the width of the central lobe of the
directed emission $w_{\rm L}$ between the first nulls at
$z=45\,a$. For the geometry considered in Ref.~\cite{moreno}, the
surface mode power is $P_{\rm S}=0.0030$, while the width of the
central lobes is $w_{\rm L} =18.1 \, a$.

A significant drawback of a surface layer design suggested in
Ref.~\cite{moreno} is a large amount of the reflected power. We
find that the reflected power can be reduced by trapping the
electrical field mostly within the surface layer, as occurs for
the uncorrugated surface. Increasing the applied wavelength by
$4.4\%$ from $\lambda = 2.45\,a$ to $\lambda=2.55\,a$ to account
for the proportionally increased distance resulting from the
corrugated surface cylinders allows us to decrease the reflected
power to $P_{\rm R}=0.048$, while increasing the directed power
and surface mode power marginally to $P_{\rm D}=0.0768$ and
$P_{\rm S}=0.0484$, respectively. A measure of the average wave
impedance in the vicinity of the waveguide shows that the
increased wavelength reduces the impedance from $\sim 1000 \Omega$
to $\sim 320 \Omega$.

In order to increase the directional power, we alter the surface
layer structure by shifting the even-numbered cylinders {\em
forward} by the distance of $\Delta z=0.4\,a$, while leaving the
odd-numbered cylinders on the lattice sites (i.e. no
displacement). As the increased distance due to this corrugation
over that of the uncorrugated distance is $7.7\%$, the applied
wavelength is increased proportionally to $\lambda=2.63865\,a$.
This new surface produces the directed power of $P_{\rm
D}=0.15418$ and decreased reflected and surface mode powers of
$P_{\rm R}=0.0318$ and $P_{\rm S}=0.0100$, respectively.
Furthermore, the central lobe of the directed emission is now
contained within $w_{\rm L} 7.79\,a$.

\begin{figure}[htbp]
\centering\includegraphics[width=9cm]{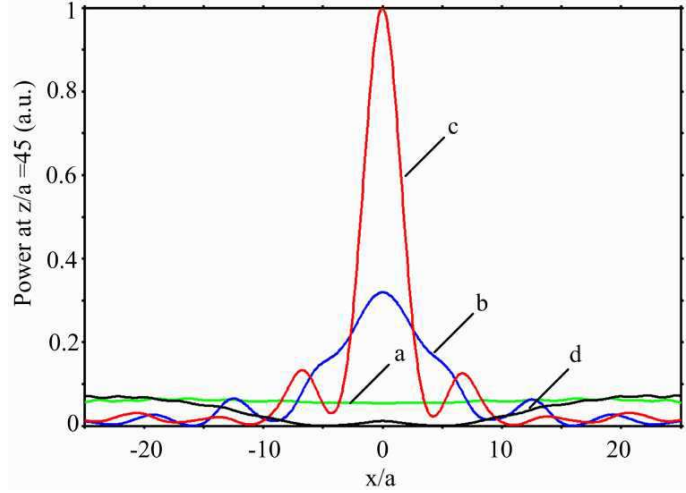}
\caption{Power density incident upon the cross-section at
$z=45\,a$ for (a) unchanged surface; (b) surface configuration
from Ref.~\cite{moreno}; (c) optimal surface configuration with
the maximized beaming, and (d) with the surface refractive index
$n=4.5$. } \label{PvZ_2_Layers}
\end{figure}

Our analysis shows that substantial improvement to the directed
power can be achieved by {\em increasing} the refractive index of
the surface layer from $n_{\rm s}=3.4$ to $n_{\rm s}=3.6$. This
results in the directed power increasing to $P_{\rm D}=0.1689$,
while decreasing the reflected and surface-mode power to $P_{\rm
R}=0.0295$ and $P_{\rm S}=0.0023$, respectively. The width of the
directed beam's central lobe resulting from the increased surface
layer's refractive index is $w_{\rm L} =9.553\,a$. The increased
power is achieved by decreasing the light-line slope, thus placing
the surface mode closer to the continuum of radiative modes.

Additional improvement of the directed power can be achieved by
decreasing the radius of the cylinders one layer prior to the
surface layer, $z=8\,a$ to $r_{\rm s-1}=0.135\,a$.  This change
induces a near-surface defect mode that leaks coherently into the
surface layer before being radiated, increasing the directed power
to $P_{\rm D}=0.2104$, the reflected power, to $P_{\rm R}=0.1028$,
and decreasing the surface power to $P_{\rm S}=0.0078$. The width
of the central lobe of the directed emission becomes $w_{\rm L}
=8.642\,a$. The spatial distribution of the Poynting vector for
this optimal design is shown in Fig.~\ref{Poynting_2_Layers}(c). A
comparison of the significantly enhanced beaming over the standard
interface and that of Ref.~\cite{moreno} is provided in
Fig.~\ref{PvZ_2_Layers}, for a cross-section of the power density
measured at $z=45\,a$.

\begin{figure}[htbp]
\centering\includegraphics[width=8cm]{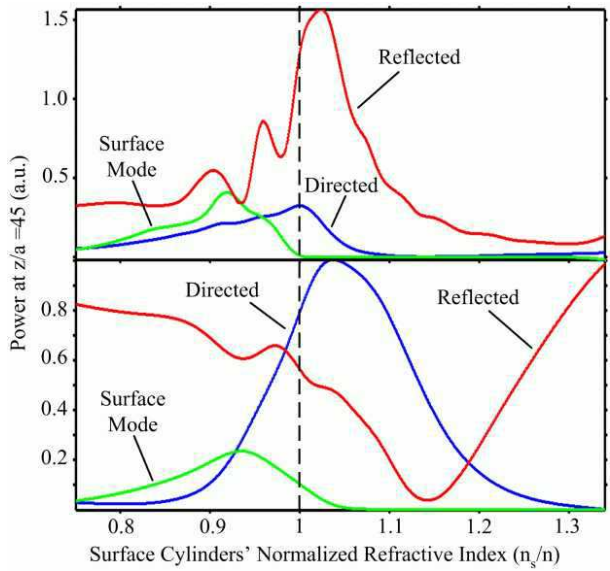}
\caption{Normalized power density incident upon a cross-sectional
length of $2\,a$ centered at $x=0$ and $z=45\,a$ as the normalized
refractive index of the surface cylinders varies. Top: surface
layer used in Ref.~\cite{moreno}. Bottom: with the surface
cylinders' radius reduced to $r_{\rm s}=0.09\,a$ and the
refractive index $n_{\rm s}=3.6$, with $N=9$ odd-numbered
cylinders displaced $\Delta z=+0.4\,a$, and the radius of the
cylinders in the layer prior to the surface layer reduced to
$r_{\rm s-1}=0.135\,a$. } \label{PvN_2_Layers}
\end{figure}

Control of the directed emission is achieved through the
manipulation of the refractive index of the surface layer
cylinders. This is illustrated in the attenuation of the directed
power shown in Fig.~\ref{Poynting_2_Layers}(d), where the
refractive index of the surface cylinders is increased to the
value $n=4.5$. In this case, the outgoing beam splits, the
directed power vanishes, and the surface-mode is in cut-off with a
localized state formed within the first two surface cylinders next
to the waveguide exit. Figure~\ref{PvZ_2_Layers}(d) shows a
cross-section of the power density measured at $z=45\,a$ for the
beam splitting depicted in Fig.~\ref{Poynting_2_Layers}(d).

\begin{figure}[htbp]
\centering\includegraphics[width=8cm]{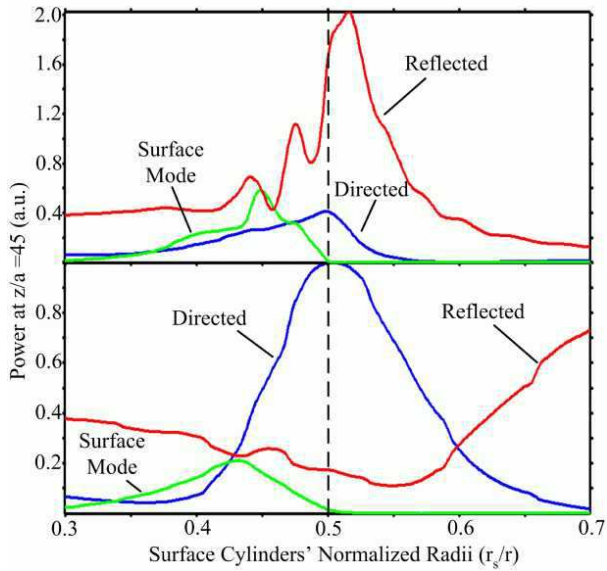}
\caption{Normalized power density incident upon a cross-section of
the length $2\,a$ centered at $x=0$ and $z=45\,a$ as the radius of
the surface cylinders varies. Top: surface layer from
Ref.~\cite{moreno} but for different values of $r_{\rm s}$.
Bottom: with the radius of the surface cylinders reduced to=20
$r_{\rm s}=0.09\,a$ and their refractive index increased to=20
$n_{\rm s}=3.6$, with $N=9$ odd-numbered cylinders displaced by
$\Delta z=+0.4\,a$.}
\label{PvR_1_Layer}
\end{figure}

The effect produced by a change of the surface refractive index is
demonstrated is Fig.~\ref{PvN_2_Layers} where the index is varied
from $n=2.4$ to $n=4.4$.  As already mentioned, the refractive
index of the surface layer has a profound effect on both the
directed and reflected powers, suggesting that it could be used
not only for achieving a control over the beaming effect but also
for matching the waveguide to the surrounding media.

The influence of the radius of the surface cylinders on the
beaming effect is summarized in Fig.~\ref{PvR_1_Layer}, where the
radius is varied from $r_{\rm s}=0.045\,a$ to $r_{\rm s}=0.2\,a$.
The radius is the key parameter in the inducement of the surface
mode and these results illustrate clearly that the optimum radius
is indeed close to $r_{\rm s}=0.9\,a$ used in Ref.~\cite{moreno}.

In conclusion, we have implemented different strategies for the
enhancement of the light beaming effect by engineering the surface
modes of photonic crystals. In particular, we have revealed that,
in comparison with the previous studies, the substantial
enhancement of the light emission and improved light beaming can
be achieved by increasing the refractive index of the surface
layer. We have provided a link of the observed enhancement of the
directional emission with the properties of the surface modes
supported by the photonic crystal interface.

We acknowledge a partial support of the Australian Research
Council and useful discussions with Sergei Mingaleev and Costas
Soukoulis.

\end{document}